\documentclass[a4paper,twocolumn,pre,showpacs]{revtex4}
\usepackage{amsmath,amssymb,amsfonts}
\usepackage{graphics}

\newcommand{\ham}{\mathcal{H}}
\newcommand{\fe}{^{\,\!}}

\begin{document}

\title{Two-dimensional lattice-fluid model with water-like anomalies}

\author{C. Buzano, E. de Stefanis, A. Pelizzola, and M. Pretti}

\affiliation{Istituto Nazionale per la Fisica della Materia (INFM)
and Dipartimento di Fisica, \\ Politecnico di Torino, Corso Duca
degli Abruzzi 24, I-10129 Torino, Italy}

\date{\today}

\begin{abstract}
We investigate a lattice-fluid model defined on a two-dimensional
triangular lattice, with the aim of reproducing qualitatively some
anomalous properties of water. Model molecules are of the
``Mercedes Benz'' type, i.e., they possess a $D_3$ (equilateral
triangle) symmetry, with three bonding arms. Bond formation
depends both on orientation and local density. We work out phase
diagrams, response functions, and stability limits for the liquid
phase, making use of a generalized first order approximation on a
triangle cluster, whose accuracy is verified, in some cases, by
Monte Carlo simulations. The phase diagram displays one ordered
(solid) phase which is less dense than the liquid one. At fixed
pressure the liquid phase response functions show the typical
anomalous behavior observed in liquid water, while, in the
supercooled region, a reentrant spinodal is observed.
\end{abstract}


\maketitle

\section{Introduction}

Water is an anomalous fluid with respect to several thermodynamic
properties~\cite{EisenbergKauzmann1969,Franks1982,Stanley2003}. At
ordinary pressures the solid phase (ice) is less dense than the
corresponding liquid, the liquid phase has a temperature of
maximum density, while both isothermal compressibility and
isobaric heat capacity display a minimum as a function of
temperature. Moreover, the heat capacity is unusually large. There
is general agreement, among physicists, that an explanation of
such anomalous properties is to be found in the peculiar features
of hydrogen bonds, and the ability of water molecules to form such
kind of bonds~\cite{Stanley1998,Poole1994}. It is also widely
believed that the same physics should be responsible of the
unusual properties of water as a solvent for apolar
compounds~\cite{FrankEvans1945,Stillinger1980}, that is of the
hydrophobic effect, of high importance in
biophysics~\cite{Dill1990}. Nevertheless, a comprehensive theory
which explains all of these phenomena has not been developed yet.
A lot of work has been done in ``realistic''
simulations~\cite{StillingerRahman1974,Jorgensen1983,MahoneyJorgensen2000,Stanley2002},
based on different interaction potentials, but they generally
require a large computational effort, and it is not always easy to
understand which detail of the model is important to determine
certain properties. On the contrary, simplified models generally
need easier numerical calculations and allow quite easily to trace
connections between microscopic interactions and macroscopic
properties~\cite{BellLavis1970,BenNaim1971,Bell1972,Lavis1973,BellSalt1976,%
LavisChristou1977,LavisChristou1979,%
MeijerKikuchiVanRoyen1982,%
HuckabyHanna1987,%
SastrySciortinoStanley1993jcp,%
RobertsDebenedetti1996,%
SilversteinHaymetDill1998}. A simplified mechanism which has been
proposed to describe the relevant physics of hydrogen bonding is
the following one (see for instance
Refs.~\onlinecite{Stanley1994,Poole1994}). Hydrogen bond formation
requires that the two involved molecules are in certain relative
orientations and stay (on average) at a distance which is larger
than the optimal distance for Van der Waals interaction. In other
words there exists a competition between Van der Waals interaction
(allowing {\em higher density} and {\em higher orientational
entropy}, but resulting in a {\em weaker bonding}) and hydrogen
bonding (requiring {\em lower density} and {\em lower
orientational entropy}, but resulting in a {\em stronger
bonding}). This simple mechanism has been implemented in different
models, both
on-~\cite{SastrySciortinoStanley1993jcp,%
RobertsDebenedetti1996,PatrykiejewPizioSokolowski1999,BruscoliniPelizzolaCasetti2002}
and off-lattice~\cite{SilversteinHaymetDill1998}, in
3~\cite{SastrySciortinoStanley1993jcp,%
RobertsDebenedetti1996} as well as 2
dimensions~\cite{SilversteinHaymetDill1998,PatrykiejewPizioSokolowski1999,BruscoliniPelizzolaCasetti2002}.
One of them is the 2-dimensional Mercedes Benz model, originally
proposed by Ben-Naim~\cite{BenNaim1971}, in which model molecules
possess three bonding arms arranged as in the Mercedes Benz logo.
In recent papers by Dill and
coworkers~\cite{SilversteinHaymetDill1998,SilversteinHaymetDill1999},
a similar (off-lattice) model has been simulated at constant
pressure by a Monte Carlo method, allowing to describe in a
qualitatively correct way several anomalous properties of liquid
water and also of hydrophobic solvation. Nevertheless, in view of
investigations on the behavior of water in contact with other
chemical species, as it happens for instance in several biological
processes, it would be desirable to obtain an even simpler
representation of the physics of hydrogen bonding.

In this paper we investigate a model of the Mercedes Benz type on
the triangular lattice, with a twofold purpose. As mentioned
above, we are first meant to explore the possibility of obtaining
a simpler model with the same underlying physical mechanism, and
with qualitatively the same macroscopic properties. Moreover, we
are interested in extending the model analysis to the global phase
diagram and in particular to the supercooled regime, in which
water anomalies are thought to find an explanation. Such a
detailed analysis is just made easier by increased simplicity.
Working on a lattice, we have to resort to a trick to describe
hydrogen bond weakening, when the two participating molecules are
too close to each other. Such a trick is similar to the one
proposed by Roberts and Debenedetti for their 3-dimensional
model~\cite{RobertsDebenedetti1996,RobertsPanagiotopoulosDebenedetti1996}.
The energy of any formed bond is increased (weakened bond) of some
fraction by the presence of a third molecule on a site close to
the bond (i.e., on the third site of the triangle). Due to the
presence of only three bonding arms, it is not possible to
distinguish between hydrogen bond donors and acceptors, but this
seems to be of minor importance to the physics of hydrogen
bonding~\cite{SilversteinHaymetDill1998}. Let us notice that the
model has the same bonding properties as the early model proposed
by Bell and Lavis~\cite{BellLavis1970}, and the same weakening
criterion as the model recently investigated by Patrykiejew and
coworkers~\cite{PatrykiejewPizioSokolowski1999,BruscoliniPelizzolaCasetti2002},
but here non-bonding orientations are added. Such a feature is
essential to describe directional selectivity of hydrogen bonds.

The paper is organized as follows. In Sec.~II we define the model
in detail and analyze its ground state. In Sec~III we introduce
the first order approximation in a cluster variational
formulation, which we employ for the analysis. Sec.~IV describes
the results and Sec.~V is devoted to some concluding remarks.

\section{Model formulation and ground state}

The model is defined on a two dimensional triangular lattice. A
lattice site can be empty or occupied by a molecule with three
equivalent bonding arms separated by $2\pi/3$ angles. Two
nearest-neighbor molecules interact with an attractive energy
$-\epsilon$ ($\epsilon > 0$) representing Van der Waals forces.
Moreover, if two arms are pointing to each other, an orientational
term $-\eta$ ($\eta > 0$) is added to mimic the formation of a
hydrogen (H) bond. Due to the lattice symmetry, a particle can
form three bonds at most and there are only 2 bonding
orientations, when the arms are aligned with the lattice, while we
assume that $w$ non-bonding configurations exist ($w$ is another
input parameter of the model). Finally, the H~bond energy is
weakened by a term $c\eta/2$ ($c \in [0,1]$) when a third molecule
is on a site near a formed bond. In the two dimensional triangular
lattice there are two such weakening sites per bond, so that a
fully weakened H~bond energy turns out to be $-(1 - c)\eta$. Let
us notice that, in the above description, H bonding is a 3-body
interaction. The hamiltonian of the system can be written as a sum
over the triangles
\begin{equation}
  \ham =
  \frac{1}{2} \sum_{\langle r,r',r'' \rangle}
  \ham_{i_r {i_r}_{'} {i_r}_{''}}
  ,
  \label{eq:ham}
\end{equation}
where $\ham_{ijk}$ is a contribution which will be referred to as
triangle hamiltonian, and $i_r,{i_r}_{'},{i_r}_{''}$ label site
configurations for the 3 vertices $r,r',r''$, respectively.
Possible configurations are empty site ($i=0$), site with a
molecule in one of the 2 bonding orientations ($i=1,2$) or in one
of the $w$ non-bonding ones ($i=3$) (see
Tab.~\ref{tab-site-conf}). The triangle hamiltonian reads
\begin{eqnarray}
  \ham_{ijk} & = & -\epsilon(n_i n_j + n_j n_k + n_k n_i)
  \\ &&
  -\eta[h_{ij}(1 - c n_k) + h_{jk}(1 - c n_i) + h_{ki}(1 - c n_j)]
  \nonumber
  ,
  \label{eq:triham}
\end{eqnarray}
where $n_i$ is an occupation variable, defined as $n_i=0$ for
$i=0$ (empty site) and $n_i=1$ otherwise (occupied site), while
$h_{ij}=1$ if the pair configuration $(i,j)$ forms a H~bond, and
$h_{ij}=0$ otherwise. Let us notice that triangle vertices are set
on three triangular sublattices, say $A,B,C$, and $i,j,k$ are
assumed to denote configurations of sites placed on $A,B,C$
sublattices respectively. Assuming also that $A,B,C$ are ordered
counterclockwise on up-pointing triangles (and then clockwise on
down-pointing triangles), we can define $h_{ij}=1$ if $i=1$ and
$j=2$ and $h_{ij}=0$ otherwise. Let us notice that both Van der
Waals ($-\epsilon n_i n_j$) and H~bond energies ($-\eta h_{ij}$),
that are 2-body terms, are split between two triangles, whence the
$1/2$~prefactor in Eq.~\eqref{eq:ham}. On the contrary the 3-body
weakening terms ($\eta h_{ij} c n_k/2$) are associated each one to
a given triangle, and the $1/2$ factor is absorbed in the
prefactor. Let us denote the triangle configuration probability by
$p_{ijk}$, and assume that the probability distribution is equal
for every triangle (no distinction between up- or down-pointing
triangles). Taking into account that there are 2 triangles per
site, we can write the following expression for the internal
energy per site of an infinite lattice
\begin{equation}
  u = \sum_{i=0}^3 \sum_{j=0}^3 \sum_{k=0}^3 w_i w_j w_k p_{ijk} \ham_{ijk},
  \label{eq:intenergy}
\end{equation}
The multiplicity for the triangle configuration $(i,j,k)$ is given
by $w_i w_j w_k$, where $w_i = w$ for $i=3$ (non-bonding
configuration) and $w_i = 1$ otherwise (bonding configuration or
vacancy).
\begin{table}
  \caption{
    Possible site configurations,
    with corresponding labels~($i$) and multiplicities~($w_i$).
  }
  \begin{ruledtabular}
  \begin{tabular}{l|cccc}
    config. & empty & \includegraphics{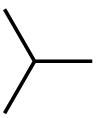} & \includegraphics{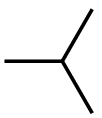} & \includegraphics{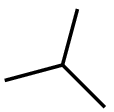}
    \cr
    \hline
    $i$ & 0 & 1 & 2 & 3 \cr
    $w_i$ & 1 & 1 & 1 & $w$ \cr
  \end{tabular}
  \end{ruledtabular}
  \label{tab-site-conf}
\end{table}

Let us now have a look at the ground state properties of the
model. In order to do so, let us investigate the zero temperature
grand-canonical free energy $\omega^\circ = u - \mu \rho$
($\mu$~being the chemical potential and $\rho$ the density, i.e.,
the average site occupation probability), which can be formally
written in the same way as the internal energy~$u$ of
Eq.~\eqref{eq:intenergy}, by replacing the triangle
hamiltonian~$\ham_{ijk}$ by
\begin{equation}
  \tilde{\ham}_{ijk} = \ham_{ijk} - \mu \frac{n_i + n_j + n_k}{3}
  \label{eq:trihamtilde}
  .
\end{equation}
We find an infinitely dilute ``gas'' phase~(G) with zero density
and zero free energy, and an ordered ``open ice''
phase~($\mathrm{I_o}$) with maximum number of H~bonds per
molecule. The latter configuration is realized through the
formation of an open (honeycomb) H~bond network with density $2/3$
and free energy
\begin{equation}
  \omega^\circ_\mathrm{I_o} = -\epsilon -\eta - 2\mu/3
  .
  \label{eq:icefreenergy}
\end{equation}
Another possibility is the ``closed ice'' phase~($\mathrm{I_c}$),
in which all interstitial sites are occupied and all hydrogen
bonds are fully weakened. The resulting free energy is
\begin{equation}
  \omega^\circ_\mathrm{I_c} = -3\epsilon -\eta(1 - c) -\mu
  .
  \label{eq:fluidfreenergy}
\end{equation}
Let us notice that it is never possible to form 3 bonds in a
triangle, which means that we have frustration. It is easy to show
that the G~phase is stable ($\omega^\circ_\mathrm{I_o}>0$) for
$\mu < \mu_\mathrm{G-I_o}$, where
\begin{equation}
  \mu_\mathrm{G-I_o} = -3(\epsilon + \eta)/2
  ,
\end{equation}
the $\mathrm{I_o}$~phase is stable ($\omega^\circ_\mathrm{I_o}<0$
and $\omega^\circ_\mathrm{I_o}<\omega^\circ_\mathrm{I_c}$) for
$\mu_\mathrm{G-I_o} < \mu < \mu_\mathrm{I_o-I_c}$, where
\begin{equation}
  \mu_\mathrm{I_o-I_c} = - 6\epsilon + 3c\eta
  ,
\end{equation}
and the $\mathrm{I_c}$~phase is stable
($\omega^\circ_\mathrm{I_c}<0$ and
$\omega^\circ_\mathrm{I_c}<\omega^\circ_\mathrm{I_o}$) for $\mu >
\mu_\mathrm{I_o-I_c}$. The $\mathrm{I_o}$ phase has actually a
stability region, i.e., $\mu_\mathrm{G-I_o} <
\mu_\mathrm{I_o-I_c}$, provided
\begin{equation}
  \eta > \frac{3}{2c+1} \epsilon
  ,
\end{equation}
which, in the worst case ($c=0$), reads $\eta > 3\epsilon$. We
shall always work in the latter regime, which is the most
significant one to describe real water properties. It is also
possible to show that, at the transition point between the open
and closed ice phases ($\mu = \mu_\mathrm{I_o-I_c}$), any
configuration built up of a honeycomb H~bond network with any
number of occupied interstitial sites has the same free energy.
Hence we expect that the $\mathrm{I_o-I_c}$ transition does not
exist at finite temperature, and actually we shall observe a
unique ice~(I) phase, in which the interstitial site occupation
probability gradually increases upon increasing the chemical
potential.

Let us finally notice that another possible phase is a homogeneous
and isotropic one in which the lattice is fully occupied and
molecules can assume only bonding configurations ($i=1,2$). This
``bonded liquid'' phase, whose free energy coincides with that of
the $\mathrm{I_c}$ phase in Eq.~\eqref{eq:fluidfreenergy}, is
observed in the $w = 0$ case, studied by Patrykiejew and
others~\cite{PatrykiejewPizioSokolowski1999,BruscoliniPelizzolaCasetti2002}.
In this scenario, non-bonding configurations are absent and the
bonded liquid ground state has, for $c \neq 1$, the same
degeneracy as the Ising triangular
antiferromagnet~\cite{BruscoliniPelizzolaCasetti2002}.
Nevertheless, in this work we shall deal with the case $w \gg 1$,
which is relevant to describe H~bond directionality. In this case
the closed ice phase is entropically favored with respect to the
bonded liquid phase, which cannot appear at finite temperature. In
conclusion, because of the introduction of non-bonding
configurations, the ground state degeneracy is removed at $T =
0^+$, where only an infinitely dilute (gas) phase and a
symmetry-broken (ice) phase are present. Such a phase behavior is
closer to the one of water than the one obtained for $w=0$.

\section{First order approximation}

We shall carry out the finite temperature analysis of the model
mainly by means of a generalized first order approximation on a
triangle cluster, which we introduce in the framework of the
cluster variation method. The cluster variation method is an
improved mean-field theory based on an approximate expression for
the entropy. In Kikuchi's original formulation~\cite{Kikuchi1951}
the entropy is obtained by an approximate counting of the number
of microstates. In a modern formulation~\cite{An1988} the
approximate entropy can be viewed as a truncation of a cluster
cumulant expansion. The truncation is justified by the expected
rapid vanishing of the cumulants upon increasing the cluster size,
namely when the cluster size becomes larger than the correlation
length of the system (the method necessarily fails near critical
points)~\cite{Morita1972}. The approximation is completely defined
by the maximum clusters left in the truncated expansion, usually
denoted as basic clusters. One obtains a free energy functional in
the cluster probability distributions, to be minimized, according
to the variational principle of statistical mechanics.

For our model we choose up-pointing triangles as basic clusters
(an analogous treatment works for down-pointing triangles). This
approximation, which seems to be good in particular for frustrated
models~\cite{NagaharaFujikiKatsura1981,Pretti2003}, is easily
shown to be equivalent to a first order approximation on a
triangle cluster~\cite{BellLavis1970}. Let us notice that the
internal energy is treated exactly, because the range of
interactions does not exceed the basic cluster size, unlike the
ordinary mean-field approximation. The grand-canonical free energy
per site $\omega = u - \mu \rho - Ts$ ($s$ being the entropy per
site), can be written as a functional in the triangle probability
distribution as
\begin{eqnarray}
  \beta \omega & = &
  \sum_{i=0}^3 \sum_{j=0}^3 \sum_{k=0}^3 w_i w_j w_k p_{ijk}
  \times \label{eq:func} \\ &&
  \left[
  \beta \tilde{\ham}_{ijk} + \ln p_{ijk}
  - \frac{2}{3} \ln \left( p^A_i  p^B_j p^C_k \right)
  \right]
  \nonumber
  ,
\end{eqnarray}
where $\beta \equiv 1/T$ (temperature is expressed in energy
units, whence entropy in natural units) and $p^X_i$ is the
probability of the $i$~configuration for a site on the
$X$~sublattice ($X=A,B,C$). The latter can be obtained as a
marginal of the triangle configuration probability~$p_{ijk}$,
namely
\begin{eqnarray}
  p^A_i & = & \sum_{j=0}^3 \sum_{k=0}^3 w_j w_k p_{ijk}
  \nonumber \\
  p^B_j & = & \sum_{i=0}^3 \sum_{k=0}^3 w_i w_k p_{ijk}
  \label{eq:marginals} \\
  p^C_k & = & \sum_{i=0}^3 \sum_{j=0}^3 w_i w_j p_{ijk}
  \nonumber
  .
\end{eqnarray}
The above expressions show that the only variational parameter
in~$\omega$ is the triangle probability distribution, that is the
64 variables $\{p_{ijk}\}$.

The minimization of~$\omega$ with respect to these variables, with
the normalization constraint
\begin{equation}
  \sum_{i=0}^3 \sum_{j=0}^3 \sum_{k=0}^3 w_i w_j w_k p_{ijk} = 1
  ,
  \label{eq:constraint}
\end{equation}
can be performed by the Lagrange multiplier method, yielding the
equations
\begin{equation}
  p_{ijk} = \xi^{-1} e^{-\beta \tilde{\ham}_{ijk}}
  {\left( p^A_i p^B_j p^C_k \right)}^{2/3},
  \label{eq:cvmeq}
\end{equation}
where $\xi$, related to the Lagrange multiplier, is obtained by
imposing the constraint Eq.~\eqref{eq:constraint}:
\begin{equation}
  \xi =
  \sum_{i=0}^3 \sum_{j=0}^3 \sum_{k=0}^3 w_i w_j w_k
  e^{-\beta \tilde{\ham}_{ijk}}
  {\left( p^A_i p^B_j p^C_k \right)}^{2/3}
  .
\end{equation}
Eq.~\eqref{eq:cvmeq} is in a fixed point form, and can be solved
numerically by simple iteration (natural iteration
method~\cite{Kikuchi1974}). In our case the numerical procedure
can be proved to lower the free energy at each
iteration~\cite{Kikuchi1974,Pretti2003}, and therefore to converge
to local minima. The solution of Eq.~\eqref{eq:cvmeq} gives the
equilibrium $\{p_{ijk}\}$ values, from which one can compute the
thermal average of every observable. Inserting these values into
Eqs.~\eqref{eq:intenergy} and \eqref{eq:func} gives respectively
the equilibrium internal energy and free energy. The latter can be
also easily expressed through the normalization constant as
\begin{equation}
  \beta \omega = -\ln \xi
  ,
\end{equation}
whence $\xi$ can be viewed as the approximate (single site)
grand-canonical partition function. It is also worth mentioning
that Eq.~\eqref{eq:cvmeq} preserves homogeneity ($p^X_i = p^Y_i$;
$\forall i,X,Y$), due to the invariance of $\tilde{\ham}_{ijk}$
under cycle permutation of the subscripts (see
Eqs.~\eqref{eq:triham} and~\eqref{eq:trihamtilde}). Let us finally
notice that the free energy expression Eq.~\eqref{eq:func} can be
also derived by considering the model on a triangular Husimi tree
(triangle cactus)~\cite{Pretti2003} as a bulk free energy density,
that is the free energy contribution far enough from the boundary,
where an invariance condition for the configuration probability of
the triangles is assumed to hold.

\section{Results}

\subsection{Phase diagrams}

\begin{figure}
  \resizebox{80mm}{!}{\includegraphics*{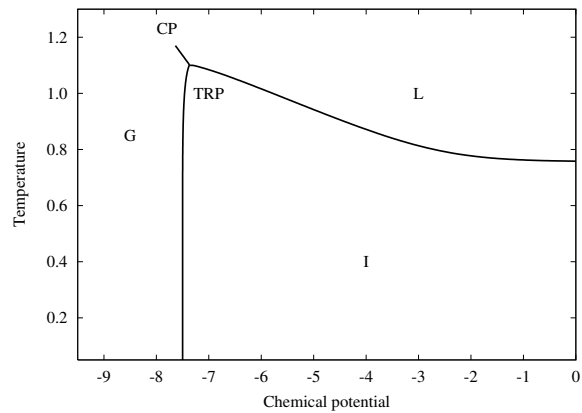}}
  \caption{
    Temperature ($T/\epsilon$) vs. chemical potential ($\mu/\epsilon$) phase diagram
    for $w = 50$, $\eta/\epsilon = 4$, $c = 0.5$.
    G, L, and I denote the gas, liquid and solid (ice) phases
    respectively. CP denotes the critical point and TRP the triple
    point.
  }
  \label{phasediag}
\end{figure}

In order to provide a first insight into the model, let us report
in Fig.~\ref{phasediag} the phase diagram in the chemical
potential-temperature plane, for $\eta/\epsilon = 4$, $c = 0.5$,
and $w = 50$. Three phases can be observed: An ice~(I) phase, with
broken symmetry among the three sublattices, a liquid~(L) phase
and a gas~(G) phase. The latter two phases preserve the sublattice
symmetry but the liquid phase has a higher density. The ice phase
has a lower density than the liquid phase, and its structure
reminds that of ground state ice, with interstitial sites occupied
by molecules in non-bonding configurations. We can observe a
triple point~(TRP), in which the three phases coexist, and a
gas-liquid critical point~(CP). All displayed transition lines are
first-order. The above phase diagram shares several properties
with the one of real water. Other crystalline phases, such as a
real close-packed ice, cannot be reproduced by the model.

\begin{figure}[b]
  \resizebox{80mm}{!}{\includegraphics*{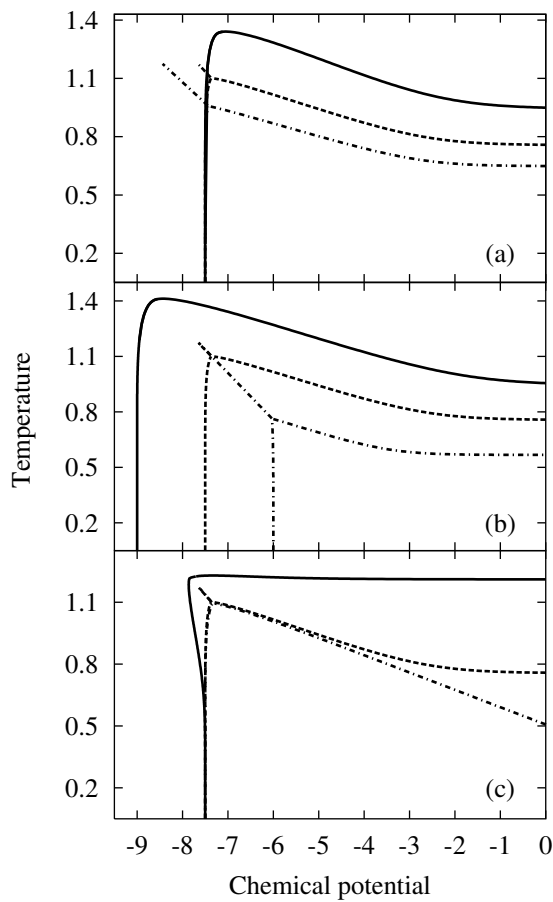}}
  \caption{
    The same phase diagram as in Fig.~\protect\ref{phasediag} (dashed lines)
    compared to different parameter choices:
    (a) $w = 20$ (solid lines) and $w = 100$ (dash-dotted lines);
    (b) $\eta/\epsilon = 5$ (solid lines) and $\eta/\epsilon = 3$ (dash-dotted lines);
    (c) $c = 0.8$ (solid lines) and $c = 0.2$ (dash-dotted lines).
  }
  \label{manyphase}
\end{figure}

Let us now investigate the role of model parameters, by analyzing
phase diagrams obtained for different values. In
Fig.~\ref{manyphase}a, $\eta/\epsilon$ and $c$ are left unchanged,
while the number of non-bonding configurations~$w$ is varied
within the interval $[20,100]$. Upon increasing~$w$, the liquid
phase turns out to be more stable with respect to the ice phase,
and the I-L transition temperature decreases. On the contrary, for
lower $w$ values, the I phase is increasingly stabilized and the
I-L transition temperature increases. For $w=20$ the whole L-G
coexistence and also the critical point disappears. Such a
behavior can be explained by the fact that the L phase is
characterized by a higher number of non-bonding molecules than the
I~phase, in which bonding molecules tend to form an ordered
structure. Therefore high $w$ values largely increase the liquid
phase entropy.

In Fig.~\ref{manyphase}b, $w$ and $c$ are held fixed and the
ratio~$\eta/\epsilon$ is varied within the interval $[3,5]$. Let
us notice that we have restricted the investigation to cases in
which the orientational (H~bond) interaction is stronger than the
non-orientational one, which is the case for real water. It turns
out that the ratio $\eta/\epsilon$ affects the stability of the
I~phase with respect to both the G and L phases. In fact higher
values of~$\eta$ means stronger H~bond, which favors the I phase,
that is the only extensively H-bonded phase. On the contrary the L
and G phases are dominated by non-oriented interactions with
coupling constants $\epsilon$, therefore both these two phases are
unfavored by high $\eta/\epsilon$~values. Even in this case the
L-G coexistence may become metastable.

The ice phase at high pressures has maximum  density and number of
weakening molecules per H bond. Raising $c$, the stability  of
this configuration is lowered with respect to the liquid phase
with few H bonds. This is shown in Fig.~\ref{manyphase}c where
$\eta/\epsilon$ and $w$ are fixed and the weakening parameter $c$
is varied in its interval of definition $[0,1]$. This trend is
reversed for low $w$~values ($w = 0$ as well), because in the
latter case the liquid has the maximum number of fully weakened
bonds.

\begin{figure}
  \resizebox{80mm}{!}{\includegraphics*{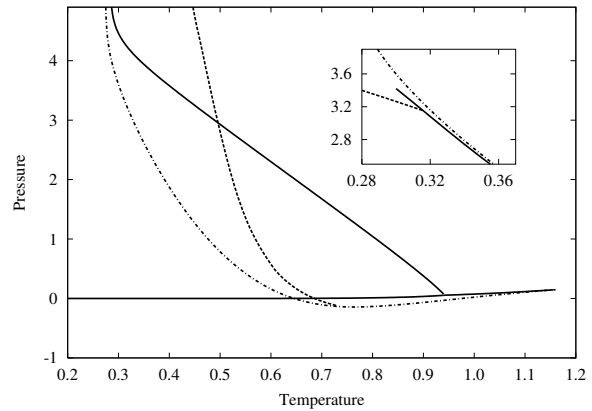}}
  \caption{
    Pressure ($P/\epsilon$) vs. temperature ($T/\epsilon$) phase diagram
    for $w = 20$, $\eta / \epsilon = 3$ and $c = 0.8$.
    Solid lines denote first order transitions,
    a dashed line denotes the TMD~locus,
    and a dash-dotted line denotes the stability limit for the liquid phase.
    The inset displays, in addition,
    the locus of divergence of the density response functions at low temperature
    (solid line) with its ``critical'' point
    and the Kauzmann line (dashed line).
  }
  \label{pT_spin_max}
\end{figure}

\begin{figure}
  \resizebox{80mm}{!}{\includegraphics*{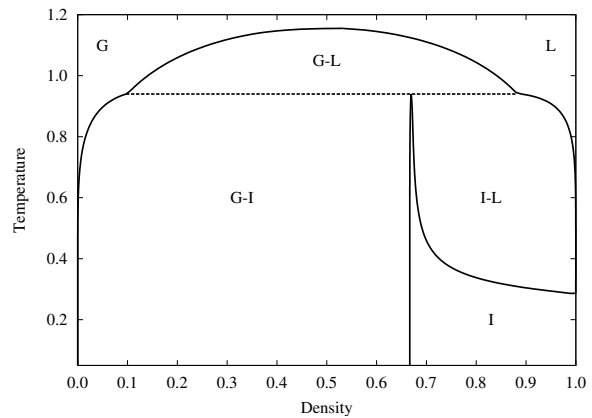}}
  \caption{
    Temperature ($T/\epsilon$) vs. density ($\rho$) phase diagram
    for $w = 20$, $\eta / \epsilon = 3$ and $c = 0.8$.
    Solid lines denote phase boundaries;
    a thin dashed line corresponds to the triple point.
    Phase labels as in Fig.~\protect\ref{phasediag};
    double labels denote two-phase coexistence regions.
  }
  \label{density}
\end{figure}

In the next part of this work we focus on a particular choice of
parameters ($w = 20$, $\eta / \epsilon = 3$ and $c = 0.8$) which,
from the above analysis, turn out to correspond to a water-like
phase diagram. Fig.~\ref{pT_spin_max} shows the
temperature-pressure phase diagram, and Fig.~\ref{density} the
temperature-density phase diagram. Let us notice that pressure~$P$
is simply given by $P = -\omega$ (the volume per site is assumed
to be equal to~1, i.e., pressure is expressed in energy units),
due to the fact that the free energy has been defined as a
grand-canonical potential.

\subsection{TMD locus and stability limits}

One of the water anomalies that the present model is able to
reproduce is the temperature of maximum density (TMD) along
isobars for the liquid phase. Joining TMD at different pressures
defines the so called TMD locus, which is a negatively sloped line
in the $T$-$P$ phase diagram of real water. We determine the
TMD~locus numerically, by adjusting the chemical potential in
order to fix the pressure and then imposing that the (isobaric)
thermal expansion coefficient vanishes.

The limit of stability of the liquid phase (spinodal) is the locus
in which the metastable liquid ceases to be a minimum of the free
energy, and becomes a saddle point. The stability limit can be
obtained by studying the eigenvalues of the hessian matrix of the
free energy~\cite{Debenedetti1996}
\begin{eqnarray}
  &&
  \frac{\partial^2 (\beta \omega)}
  {\partial p_{i\fe j\fe k\fe} \partial p_{i'j'k'}}
  = w_{i}w_{j}w_{k} \biggr\{ \frac{\delta_{ii'}\delta_{jj'} \delta_{kk'}}{p_{ijk}}
  \label{eq:hessian} \\ &&
  - \frac{2}{3} \biggr[
  \frac{\delta_{ii'} w_{j'} w_{k'}}{p^A_i} +
  \frac{w_{i'} \delta_{jj'} w_{k'}}{p^B_j} +
  \frac{w_{i'} w_{j'} \delta_{kk'}}{p^C_k}
  \biggr]\biggr\}
  \nonumber
  .
\end{eqnarray}
Let us notice that, when the liquid phase stability is lost (some
eigenvalue of the above matrix vanishes), also the corresponding
fixed point of the natural iteration equations~\eqref{eq:cvmeq}
becomes unstable. In order to determine the stability limit with
respect to the symmetry-broken ice phase, it is sufficient to
impose homogeneity during the iterative procedure, which is done
by replacing Eqs.~\eqref{eq:marginals} with
\begin{equation}
  p^A_i = p^B_i = p^C_i =
  \sum_{j=0}^3 \sum_{k=0}^3 w_j w_k
  \frac{p_{ijk} + p_{kij} + p_{jki}}{3}
  .
  \label{eq:homogeneity}
\end{equation}
This trick cannot be applied when the liquid stability is lost
with respect to a homogeneous phase, because the liquid fixed
point of equations~\eqref{eq:cvmeq} becomes definitely unstable,
due to divergence of the density response functions. In the latter
case the spinodal is determined by solving the eigenvalue problem
for the hessian matrix rewritten by forcing the homogeneity
condition~\eqref{eq:homogeneity}.

The results are shown in Fig.~\ref{pT_spin_max}. The stability
limit of the liquid with respect to the gas phase starts from the
critical point and reaches a minimum in the negative pressure
region. After this point the line becomes negatively sloped and
joins continuously the stability limit with respect to the ordered
ice phase. The TMD~locus intersects the limit of stability in its
minimum in the $T$-$P$ plane, according to the predictions of
Speedy and
Debenedetti~\cite{Speedy1982I,Speedy1982II,Speedy1987,%
DebenedettiDantonio1986I,DebenedettiDantonio1986II,DantonioDebenedetti1987,DebenedettiDantonio1988},
based on thermodynamic consistency arguments. In fact the
TMD~locus causes the liquid limit of stability line to retrace,
giving rise to a tensile strength maximum and to a continuous
boundary. Let us recall that, while at the stability limit with
respect to the gas phase, the density response functions diverge,
this is not the case at the stability limit with respect to the
ordered phase. Nevertheless we can observe that the density
response functions tend to diverge also upon decreasing
temperature, as observed experimentally. The locus of divergence,
terminating at some kind of critical point, can be defined, in the
framework of a simplified variational free energy forced to
describe a homogeneous system, as an additional stability limit
with respect to a low density liquid phase. Such ``phase''
corresponds to a saddle point of the original (not symmetrized)
free energy, unstable with respect to the solid phase. As the low
pressure solid phase reminds the ground state ``open ice''
structure, which is three-fold degenerate, the triangle
probability distribution of the low density liquid phase turns out
to be essentially an arithmetic average over the three ice
distributions. The unphysical nature of this solution is also
reflected in its negative entropy. The divergence locus, together
with the locus at which the liquid phase entropy vanishes
(Kauzmann line), are shown for completeness in the inset of
Fig.~\ref{pT_spin_max}. Upon increasing temperature the divergence
locus meets the spinodal tangentially and they become the same
curve ending in the ``true'' gas-liquid critical point.

\subsection{Response functions}

\begin{figure}
  \resizebox{80mm}{!}{\includegraphics*{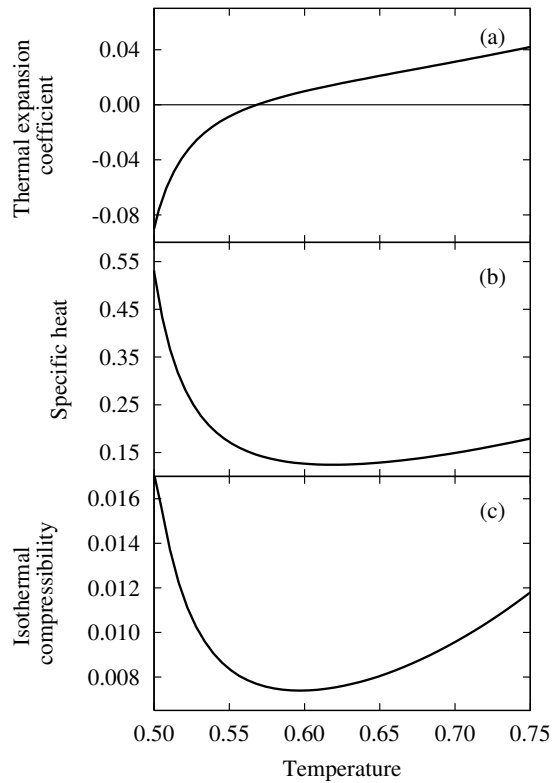}}
  \caption{
    Response functions at constant pressure ($P/\epsilon = 1$)
    as a function of temperature ($T/\epsilon$):
    (a) thermal expansion coefficient ($\epsilon \alpha_P$);
    (b) specific heat ($c_P$);
    (c) isothermal compressibility ($\epsilon \kappa_T$).
  }
  \label{response}
\end{figure}
Let us now investigate the density response functions and the
specific heat of the liquid at constant pressure $P/\epsilon = 1$
(pressure is kept fixed by numerically adjusting the chemical
potential~$\mu$). It turns out that these functions display an
anomalous behavior similar to that of real liquid water. The first
response function we consider is the thermal expansion coefficient
$\alpha_P = (-\partial\ln\rho/\partial T)_P$, which is
proportional to the entropy-specific volume cross-correlation. For
a typical fluid $\alpha_P$~is always positive because if in a
region of the system the specific volume is a little larger then
the average, then the local entropy is also larger, i.e., the two
quantities are positively correlated. On the contrary, for our
model $\alpha_P$ (Fig.~\ref{response}a) displays an anomalous
behavior. As temperature is lowered $\alpha_P$ vanishes (at
the~TMD), becomes negative, and finally tends to diverge. As
previously mentioned, divergence can be observed only for pressure
values less than some ``critical'' pressure. Anyway, before
divergence is actually reached, the liquid loses stability with
respect to the ice phase.

The trend of the isothermal compressibility $\kappa_T =
(\partial\ln\rho/\partial P)_T$ is also anomalous
(Fig.~\ref{response}c). For a typical liquid $\kappa_T$ decreases
as one lowers temperature, because it is proportional to density
fluctuations, which decrease upon decreasing temperature. On the
contrary, in Fig.~\ref{response}c we can observe that $\kappa_T$,
once reached a minimum, begins to increase upon decreasing
temperature. Such a behavior is observed in real liquid water. An
analogous behavior characterizes the constant pressure specific
heat $c_P = (-T
\partial^2 \mu/\partial T^2)_P$ (Fig.~\ref{response}b).

\subsection{Numerical simulation}

We have studied the model in the first order approximation to
obtain easily detailed information about phase diagrams and in
particular the metastable region. In order to check this
approximation and obtain an estimate of its quantitative accuracy,
we have also performed some (grand-canonical) Monte Carlo
simulations on a $60 \times 60$ triangular lattice with periodic
boundary conditions. From the very beginning, we have chosen quite
a low number of non-bonding configurations for our analysis
($w=20$), in order to increase the speed of simulation dynamics.
In fact a lower $w$ value corresponds to a smaller configuration
space. We report some results in the following.

\begin{figure}
  \resizebox{80mm}{!}{\includegraphics*{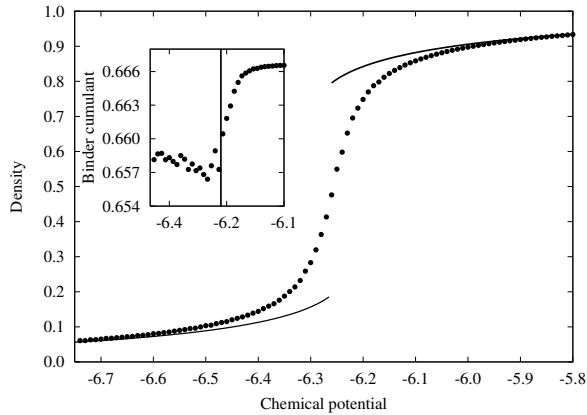}}
  \caption{
    Gas-liquid transition at fixed temperature ($T/\epsilon = 1.05$),
    upon varying the chemical potential ($\mu/\epsilon$):
    First order approximation results (solid line)
    compared to Monte Carlo simulations (scatters),
    for $w = 20$, $\eta/\epsilon = 3$, and $c = 0.8$.
    The inset displays the Binder cumulant minimum,
    together with the transition point predicted by
    the first order approximation (vertical line).
  }
  \label{transition}
\end{figure}
In Fig.~\ref{transition} we show a first order transition between
the gas and the liquid phases along a constant temperature path
$T/\epsilon = 1.05$, quite less than the critical temperature. At
the critical point the correlation length increases and the
approximation may give worse predictions. Fig.~\ref{transition}
suggests that the first order approximation well localizes the
transition and that far enough from the critical point its
predictions are nearly quantitative. Of course Monte Carlo
simulations display smooth density variations, due to finite size
effects, but the Binder cumulant (inset), displaying a minimum,
gives evidence of a first order transition.

\begin{figure}
  \resizebox{80mm}{!}{\includegraphics*{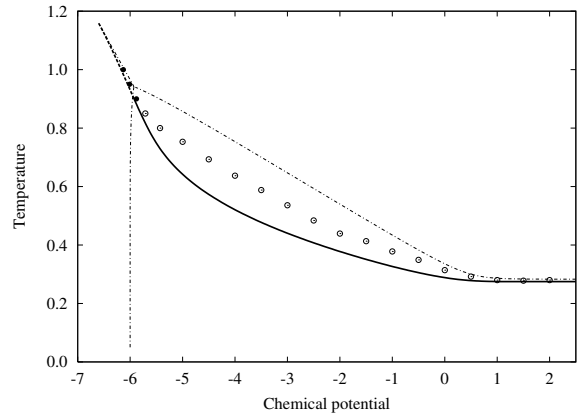}}
  \caption{
    Stability limits from first order approximation (thick lines)
    and homogeneous nucleation points from Monte Carlo simulations
    (scatters), for $w = 20$, $\eta/\epsilon = 3$, and $c = 0.8$.
    Open circles and a solid line denote the stability limit to the ice phase,
    filled circles and a dashed line the stability limit to the gas phase.
    Thin dash-dotted lines denote equilibrium phase boundaries.
  }
  \label{MCspin}
\end{figure}
The reentrance of the liquid stability limit, which is one of the
striking features of the (metastable) phase diagram of this model,
is also confirmed by simulations. Performing simulations in the
metastable region, the spinodal has been determined by an
arbitrary criterion for the life time of the metastable phase (100
Monte Carlo steps), as it has been done in previous
studies~\cite{SastrySciortinoStanley1993jcp}. Such a criterion
allows us to find the kinetically controlled limit of supercooling
(homogeneous nucleation locus), shown in Fig.~\ref{MCspin}, along
with the corresponding first order approximation result. Both
methods show a reentrant spinodal forming a continuous boundary.
The simulations also confirm the distinction between liquid limit
of stability with respect to the gas or to the ice phase, as in
the first order approximation.

\section{Discussion and conclusions}

In this paper we have investigated a 2-dimensional lattice model
in which model molecules possess three equivalent bonding arms,
and bonding energy depends on the presence of neighbor molecules,
giving rise to a 3-particle interaction. The observed behavior is
qualitatively similar to that of water, exhibiting the correct
anomalies. Upon supercooling, $\kappa_T$ and $c_P$ increase and
$\alpha_P$ becomes negative and large in magnitude. Nevertheless,
at ordinary pressures (less than the critical pressure) the
density anomaly ($\alpha_P = 0$) is found in the metastable liquid
region. We have also determined the spinodal limits to the liquid
state, and pointed out the relationship between these limits and
the TMD~locus. The growth in the response functions upon
decreasing temperature can be interpreted on the basis of a
reentrant spinodal scenario. The liquid-gas spinodal meets the TMD
locus at the reentrance point, as required by thermodynamic
consistency. Actually the reentrant spinodal conjecture is one of
the possible theoretical explanations of water anomalies, and some
experimental results are consistent with this
explanation~\cite{ZhengDurbenWolfAngell1991}. Nevertheless, it is
important to note that, for the specific case of water,
alternative interpretations of the stability problem exist, based
on the second critical point conjecture~\cite{Stanley1998}. The
latter, supported by molecular dynamics
simulations~\cite{Stanley2002}, seems to be more consistent with
the existence, in the negative pressure region, of a monotonic
liquid-gas spinodal and a reentrant TMD~locus. On the contrary,
our model displays a metastable liquid state which is bounded by a
spinodal both at positive as well as negative pressures, forming a
continuous boundary. The lower temperature part of the boundary is
the limit of stability with respect to the ordered ice phase,
while the higher temperature part is the limit of stability with
respect to the gas phase. While the response functions diverge at
the liquid-gas spinodal, at the liquid-solid spinodal they do not,
even if they tend to higher values. Anyway, in our framework, it
is also possible to investigate the behavior of the unstable
liquid (a saddle point of the variational free energy) and
determine the locus of divergence. The latter always turns out to
lie at a temperature less than the limit of stability, according
to experiments~\cite{SpeedyAngell1976}. It also turns out that the
divergence locus terminates at some kind of critical point,
meaning that response functions should not show divergent-like
behavior for pressure values greater than some critical pressure.

Let us notice that a previous lattice model on the 3-dimensional
body centered cubic lattice had pointed out a qualitatively
similar behavior~\cite{SastrySciortinoStanley1993jcp}.
Nevertheless, in such a model, orientational degrees of freedom of
water are not treated explicitly and two equivalent sublattices
are artificially distinguished by the hamiltonian. This is
necessary to favor an open structured phase. Moreover, the
analytical treatment is based on the determination of a
temperature dependent 2-particle interaction. On the contrary in
our model there exists an explicit, though simplified, modelling
of hydrogen bonding and no temperature dependent interaction is
introduced. The open structured phase is favored in principle by
the triangular lattice structure.

We have mentioned in the Introduction that the present model is
actually an extension over an early model proposed by Bell and
Lavis~\cite{BellLavis1970} (corresponding to the case in which
$w=0$ and $c=0$) and over a recent model investigated by
Patrykiejew and
coworkers~\cite{PatrykiejewPizioSokolowski1999,BruscoliniPelizzolaCasetti2002}
(corresponding to $w=0$). The former model in the same
approximation actually displays, for $\eta/\epsilon > 3$, a
density anomaly (without singularities), but we have verified that
the anomaly occurs in a negative entropy region. The latter model
shows an unrealistic phase diagram, in which, for high enough
pressure, the liquid phase extends its stability region down to
zero temperature. In the present work we have shown that the
addition of non-bonding configurations to such a simple class of
2-dimensional lattice models allows us to reproduce a
qualitatively correct water-like behavior. Moreover, this result
has been obtained in a computationally much simpler way than a
conceptually similar model with continuous degrees of freedom,
that is the Mercedes-Benz one. The latter model is highly
appealing, because of its ability to explain most phenomena
related to hydrophobicity~\cite{SilversteinHaymetDill1999}.
Therefore it would be interesting to analyze also the properties
of the present model for a solution of an inert (apolar) solute,
whose peculiar properties are thought to be strictly related to
hydrogen bonding. This goes beyond the scope of the present paper
and will be the subject of a forthcoming article.


\end{document}